\documentclass[journal=nalefd,manuscript=article]{achemso}
\usepackage{chemformula} % Formula subscripts using \ch{}
\usepackage{amsmath}
\usepackage{dcolumn}% Align table columns on decimal point
\usepackage{bm}% bold math

\title{Tunable strong coupling of mechanical resonance between spatially separated \ch{FePS3} nanodrums}

\author{Makars \v{S}i\v{s}kins}
\affiliation{These authors contributed equally.}
\alsoaffiliation{%
 Kavli Institute of Nanoscience, Delft University of Technology, Lorentzweg 1, 2628 CJ, Delft, The Netherlands
}
\email{m.siskins-1@tudelft.nl}
\author{Ekaterina Sokolovskaya}%
\affiliation{These authors contributed equally.}
\alsoaffiliation{%
 Kavli Institute of Nanoscience, Delft University of Technology, Lorentzweg 1, 2628 CJ, Delft, The Netherlands
}

\author{Martin Lee}%
\affiliation{%
 Kavli Institute of Nanoscience, Delft University of Technology, Lorentzweg 1, 2628 CJ, Delft, The Netherlands
}

\author{Samuel Ma\~{n}as-Valero}%
\affiliation{%
Instituto de Ciencia Molecular (ICMol), Universitat de Val\`{e}ncia, c/Catedr\'{a}tico Jos\'{e} Beltr\'{a}n 2, 46980 Paterna, Spain
}%
\author{Dejan Davidovikj}
\affiliation{%
	Kavli Institute of Nanoscience, Delft University of Technology, Lorentzweg 1, 2628 CJ, Delft, The Netherlands
}%
\author{Herre S. J. van der Zant}%
\affiliation{%
	Kavli Institute of Nanoscience, Delft University of Technology, Lorentzweg 1, 2628 CJ, Delft, The Netherlands
}%
\author{Peter G. Steeneken}
\affiliation{%
	Kavli Institute of Nanoscience, Delft University of Technology, Lorentzweg 1, 2628 CJ, Delft, The Netherlands
}%
\alsoaffiliation{
	Department of Precision and Microsystems Engineering, Delft University of Technology,
	Mekelweg 2, 2628 CD, Delft, The Netherlands}
\email{p.g.steeneken@tudelft.nl}
\date{\today}
%Use showkeys class option if keyword
\begin{document}
\begin{abstract}
Coupled nanomechanical resonators made of two-dimensional materials are promising for processing information with mechanical modes. However, the challenge for these types of systems is to control the coupling. Here, we demonstrate strong coupling of motion between two suspended membranes of the magnetic 2D material \ch{FePS3}. We describe a tunable electromechanical mechanism for control over both the resonance frequency and the coupling strength using a gate voltage electrode under each membrane. We show that the coupling can be utilized for transferring data from one drum to the other by amplitude modulation. Finally, we also study the temperature dependence of the coupling, and in particular how it is affected by the antiferromagnetic phase transition characteristic of this material. The presented electrical coupling of resonant magnetic 2D membranes holds promise of transferring mechanical energy over a distance at low electrical power, thus enabling novel data readout and information processing technologies.
\end{abstract}

\maketitle

\section{INTRODUCTION}
Nano-electromechanical systems (NEMS) are attracting attention of the scientific community for their potential to study novel quantum and electromagnetic effects at the nanoscale \cite{NEMSEkinci2005,Aspelmeyer2012,MainAspelmeyer2014}. Along with that, micro- and nanoresonators have been studied for various applications, including sensitive mass detection \cite{Mass1Spletzer2008,Mass2GilSantos2009}, bandpass filters with variable properties \cite{FilterBannon2000}, logic gates \cite{LogicNEMSTsai2008,BitLogicMahboob2008, LogicMasmanidis2007} and signal amplifiers \cite{AmplifierKarabalin2011}. For instance, arrays of coupled highly-cooperative NEMS and oscillators are already utilized for the coherent manipulation of phonon populations \cite{NatPhysOkamoto2013, NEMSPhononsFaust2013, CarbonPhononsZhu2017}, and for data processing and storage \cite{MainAspelmeyer2014, NetworkShim2007}. Recently, NEMS made out of two-dimensional (2D) materials have also gained interest due to perspectives for realizing high-performance oscillators \cite{Chen2009,Chen2013} and novel sensor concepts \cite{GrapheneSensorsLemme2020}. This is not only due to the atomic thinness of these devices, but also because the fabrication methodology allows the integration of a range of materials and their heterostructures, with a wide range of magnetic, optical and electrical properties, on the same chip \cite{MishchenkoNovoselov2016}. Hence, coupling NEMS resonators made of 2D materials and heterostructures promises even more interesting implementation possibilities.

Various studies have reported mechanical coupling between different resonance modes of the same 2D membrane by mechanical, optical and electronic means \cite{GrapheneMathew2016,Liu2015,MoS2Prasad2019, Kekekler2021}. However, realisation of coupling between resonances of spatially separated 2D membranes has appeared to be more difficult and was only recently achieved via a mechanical phononic transduction mechanism \cite{GrapheneDistantLuo2018, PhononManipZhang2020}. A mechanically mediated coupling mechanism via a phonon bath was demonstrated \cite{GrapheneDistantLuo2018, PhononManipZhang2020}, but although the coupling could be adjusted via the individual resonance frequencies of the resonators, the coupling strength itself is fixed by the mechanical geometry of the structure that determined the phonon bath. In order to achieve full control over the coupling, a tunable transduction mechanism is needed which not only adjusts the degree of coupling between the resonators, but also regulates their resonance frequencies.

Here, we demonstrate an electrical transduction mechanism for coupling mechanical resonances of two spatially separated membranes made of van der Waals material, allowing control over both the resonance frequency and the coupling strength via gate electrodes. We use the mechanism to strongly couple the fundamental mechanical modes of two suspended circular antiferromagnetic \ch{FePS3} membranes that are separated by an edge-to-edge distance of $2$ microns. We show that the coupling mechanism can be utilized for transferring data from one drum to the other, a feature that is useful in data processing and storage systems. Coupling of magnetic materials like \ch{FePS3} is of interest, since their mechanical resonances can be sensitive to both the magnetic phase \cite{Siskins2020} and the magnetisation of that phase \cite{FaiMakJiang2020}. To investigate this, we study the temperature dependence of the coupling strength, and in particular how it is affected by the antiferromagnetic phase transition at the N\'{e}el temperature, $T_{\rm{N}}\sim114$ K \cite{CvTakano2004, Lee2016}. 

\section{RESULTS AND DISCUSSION}
\begin{figure*}
	\includegraphics[scale=1]{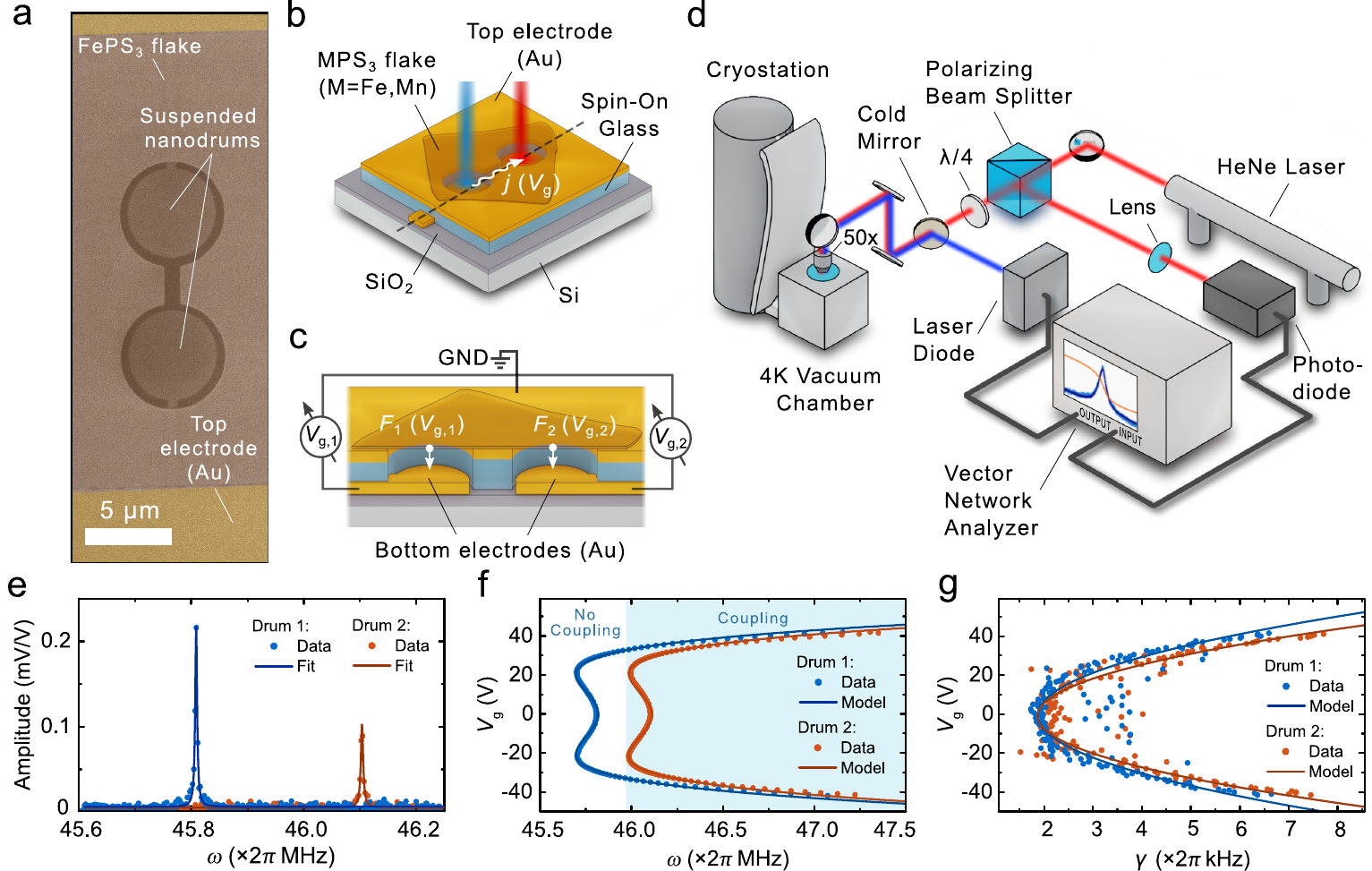}
	\caption{Measurement principle and setup. (a) False-coloured SEM image of the sample. Flake thickness: $25.6\pm0.4$ nm. (b) Schematics of device and optical measurement principle; $j(V_{\rm g})$ is the voltage-dependent coupling parameter. (c) Schematics of cross-section of the device, electrically induced force $F_{1,2}$ and gate voltage $V_{\rm{g}}$. (d) Laser interferometry setup. Red laser:  $ \lambda_{\mathrm{red}} = 632 $ nm. Blue laser: $\lambda_{\mathrm{blue}} = 405$ nm. (e) Detected resonance peaks for the two suspended drums. Filled dots - measured data, Solid lines - linear harmonic oscillator fits. (f) Resonance frequencies $\omega(V_{\rm g})$ of drums 1 and 2, extracted from fits similar to (e). Filled dots - measured data, Solid lines - continuum mechanics model \cite{ThesisChen2013,WWeber2014} (see SI 1). Blue region indicates the parameter-space where $\omega_1=\omega_2$ condition can be met. (g) Dissipation rate $\gamma(V_{\rm g})$ for two drums extracted from fits similar as shown in (e). Filled dots - measured data; Solid lines - Joule dissipation model \cite{DispCurrSong2011,GorbachevWill2017,Morell2016} (see SI 2).}
	\label{fgr:first}
\end{figure*}

\paragraph{Laser Interferometry of \ch{FePS3} Resonators.}
We fabricate two circular suspended \ch{FePS3} resonators on top of a \ch{Si}/\ch{SiO2} substrate on which we define an array of Au bottom gate electrodes. A layer of spin-on glass (SOG) \cite{Davidovikj2017} is used to electrically insulate the bottom electrodes from the top electrode as indicated in Fig.~\ref{fgr:first} (see also Methods). Then, two spatially separated circular cavities of $r=3$ $\mu$m in radius are etched in the SOG/Au top layer \cite{Davidovikj2017}, such that the local circular gate electrode with radius of $r_{\rm g}=2.5$ $\mu$m is located at the bottom of the cavity. We transfer a flake of few-layer \ch{FePS3} exfoliated from a synthetically grown bulk crystal \cite{Siskins2020} over these cavities by a dry transfer technique (see Methods) to create two separated circular membranes of the same flake as depicted in Fig.~\ref{fgr:first}a-c. As shown in Fig.~\ref{fgr:first}b, we focus red and blue lasers on these drums to excite the motion of one and measure the displacement of the other using a laser interferometry technique (see Methods), thus probing the coupling between fundamental vibration modes of these membranes. To realize the study as a function of gate voltage and corresponding electrostatically induced strain for both drums, we use local electrodes that allow us to individually adjust the gate voltage, $V_{\rm g}$, for each of the resonators (see Fig.~\ref{fgr:first}c). The single \ch{FePS3} flake, out of which the resonators are formed, is contacted via a top metal electrode to ground. A false-coloured scanning electron microscopy (SEM) image in Fig.~\ref{fgr:first}a shows a $25.6\pm0.4$ nm FePS$_3$ flake suspended over the electrodes forming two separated membrane resonators. The resonators are placed in a dry cryostat with optical access that is connected to a laser interferometry setup, as shown in Fig.~\ref{fgr:first}d (see Methods).

When operating the laser interferomery technique with both lasers focused at the same position on the same membrane, we independently characterize the resonance spectra of the fundamental membrane modes of drums 1 and 2 at a temperature of $4$ K as shown in Fig.~\ref{fgr:first}e. We fit these spectra to a harmonic oscillator model and extract the resonance frequencies $\omega_{1,2}$ as a function of $V_{\rm g}$, which are displayed with filled blue and orange dots in Fig.~\ref{fgr:first}f. The resonances $\omega_{1,2}(V_{\rm g})$ of both drums closely follow the continuum mechanics model \cite{ThesisChen2013,WWeber2014}, as shown by the solid blue and orange lines in Fig.~\ref{fgr:first}f (see SI 1). At certain values of $V_{\rm{g},1}$ and $V_{\rm{g},2}$, the frequencies $\omega_1$ and $\omega_2$ of the corresponding resonance peaks match at $\omega_1=\omega_2$, as indicated by the light blue region in Fig.~\ref{fgr:first}f. In this regime we can expect an avoided frequency crossing if the exchange of the excitation energy between the drums is sufficiently large and the drums are strongly coupled \cite{MainAspelmeyer2014}. The coupling strength is also related to the dissipation $\frac{1}{Q_{1,2}}$ of the resonators involved \cite{NatPhysOkamoto2013, GrapheneMathew2016}. Hence, we plot the corresponding mechanical energy dissipation rates $\gamma_{1,2}=\frac{\omega_{1,2}}{Q_{1,2}}$ of the \ch{FePS3} membranes in Fig.~\ref{fgr:first}g. Measured $\gamma_{1,2}(V_{\rm g})$ follow a parabolic behavior, in accordance with a Joule dissipation model (solid blue and orange lines, see SI 2), which can be attributed to capacitive displacement currents in the suspended region of the flake \cite{DispCurrSong2011,GorbachevWill2017,Morell2016}.

\paragraph{Electromechanical Coupling Model.}

\begin{figure}
	\includegraphics[scale=1]{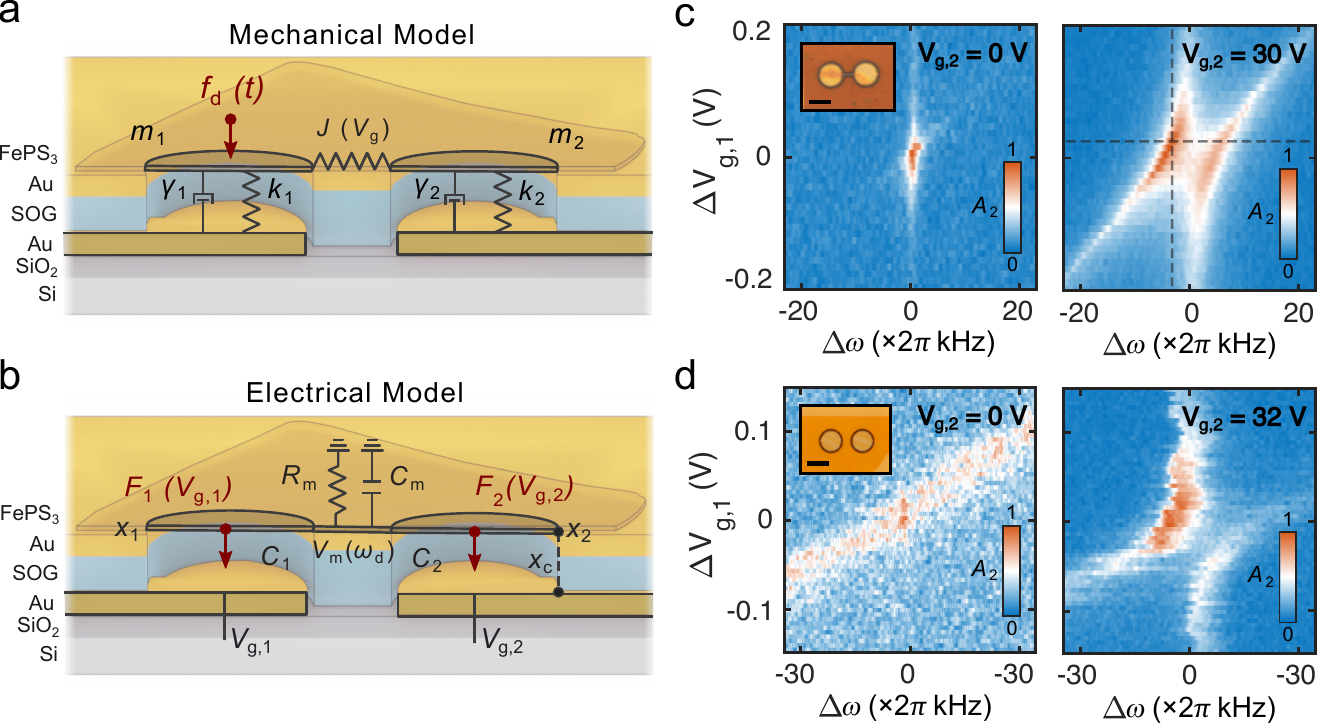}
	\caption{Strong coupling of spatially separated FePS$_3$ membrane resonators. Schematics of coupled membrane oscillators: (a) Mechanical model: $m_{1,2}$ is the effective mass, $k_{1,2}$ the effective stiffness, and $J(V_{\rm g})$ the gate voltage dependent coupling parameter. (b) Electrical model: $C_{1,2}$ is the capacitance of each drum towards the gate electrode that is kept at a voltage $V_{\rm g}$, $R_{\rm m}$ the resistance to ground, $C_{\rm m}$ the capacitance to ground, and $V_{\rm m}(\omega_{\rm d})$ the voltage between the membranes. (c) Sample with a suspended channel between the drums. Left panel: Weak coupling of motion between spatially separated drums at $V_{\rm{g},1}=36.9$~V and $V_{\rm{g},2}=0$~V and $\Delta \omega=\omega_{\rm d}-\omega_2$. Inset: Optical image of the sample. Thickness: $25.6\pm0.4$ nm. Scale bar: $6$ $\mu$m. Right panel: Strong coupling of motion between spatially separated drums at $V_{\rm{g},1}=37.2$~V and $V_{\rm{g},2}=30$~V. (d) Sample without a suspended channel between the drums (see SI 7). Left panel: Weak coupling of motion at $V_{\rm{g},1}=32.4$~V and $V_{\rm{g},2}=0$~V. Inset: Optical image of the sample. Thickness: $14.5\pm0.3$ nm. Scale bar: $6$ $\mu$m. Right panel: Strong coupling of motion at $V_{\rm{g},1}=34.5$~V and $V_{\rm{g},2}=32$~V.}
	\label{fgr:second}
\end{figure}

The mechanical behavior of coupled membrane resonators can be modeled by two coupled resonators, schematically depicted in Fig.~\ref{fgr:second}a. The motion of coupled resonators is described by: 
\begin{equation}
    \left\{
        \begin{array}{ll}
           \ddot{x}_1+\gamma_1\dot{x}_1+\omega_1^2x_1=jx_2+f_{\rm{d}}\cos{\omega_{\rm d} t}\\
           \ddot{x}_2+\gamma_2\dot{x}_2+\omega_2^2x_2=jx_1
     \end{array}
    \right. ,
    \label{eq:system_j}
\end{equation}
where $x_{1,2}$ are the membrane displacements and $f_{\rm{d}}$ is the force at a drive frequency $\omega_{\rm d}$. The coupling parameter $j=\frac{J}{\sqrt{m_1m_2}}$, where $m_{1,2}$ is the effective mass, is responsible for the transfer of energy between the two resonators and thus coupling of the mechanical motion. Several coupling mechanisms can contribute to $J$ (see SI 3). In this work, we present evidence for an electromechanical coupling mechanism for adjusting the coupling strength between two 2D material resonators.

Figure~\ref{fgr:second}b shows the schematic of the electrical circuit that mediates the coupling (see SI 4). The suspended part of the thin \ch{FePS3} flake, that covers the two cavities, is both resistively and capacitively connected to ground via the interface between the flake and the Au top electrode. We assume that the voltage $V_{\rm{m,DC}}$ that is established between drum 1 to drum 2 is zero since the Au top electrode effectively shunts potential differences between the drums. However, since the capacitance $C_{\rm m}$ between the \ch{FePS3} flake and Au top electrode is large, it dominates the electrical coupling of the flake to ground ($\frac{1}{\omega_{\rm d} C_{\rm m}}\ll R_{\rm m}$, where $R_{\rm m}$ is the resistance to ground). As is outlined in SI 4, the optothermal drive of the first drum at non-zero $V_{\rm{g,1}}$ then results in a non-zero flake voltage $V_{\rm{m,AC}}$ that causes an electrostatic force on the second drum, $F_{\rm{2,AC}}=-J_{\rm{el}}x_{10}\sin{\left(\omega_{\rm d}t\right)}$, where $x_{10}$ is the amplitude of periodic displacement and the electrical coupling parameter $J_{\rm{el}}$ is given by: 
\begin{equation}
    J_{\rm{el}}\cong\frac{\left(\varepsilon_0\pi r_{\rm g}^2\right)^2}{C_{\rm m}} \frac{V_{\rm{g, 1}}\,V_{\rm{g, 2}}}{(x_{\rm c}-x_{\rm{g, 1}})^2\,(x_{\rm c}-x_{\rm{g, 2}})^2},
    \label{eq:F2ac}
\end{equation}
where $\varepsilon_0$ is the dielectric permitivity of vacuum, $x_{\rm g}(V_{\rm g})$ the static deflection at the centre of membrane (see SI 1) and $x_{\rm c}$ the separation between the membrane and the bottom electrode. By combining equations~\ref{eq:system_j} and \ref{eq:F2ac}, it is seen that $J_{\rm{el}}$ results in a transfer of mechanical energy via electromechanical coupling between the spatially separated \ch{FePS3} drums. It is notable that the driving force on the second drum $F_{\rm{2,AC}}$ is proportional to the product of the individual gate voltages applied to each of the drums. This means that at $V_{\rm g}=0$~V on either drum, the electrical coupling parameter $J_{\rm{el}}=0$ even if $\omega_1=\omega_2$ and $f_{\rm{d}}> 0$. This property distinguishes the expected behavior of this mechanism from phonon- or tension-mediated coupling \cite{GrapheneMathew2016, MoS2Prasad2019, GrapheneDistantLuo2018, PhononManipZhang2020}, where frequency matching and a non-zero driving force acting on one of the drums are sufficient conditions for coupling and thus splitting of the resonance frequency to occur. 

We use this characteristic to provide evidence for the proposed mechanism, by first matching the resonance frequencies of the drums by tuning $\omega_{1,2}$ such that $\omega_1=\omega_2$ using electrostatic pulling, as shown in Fig.~\ref{fgr:first}f. Then we alter $V_{\rm{g},1}$ of the drum that we drive with the modulated blue laser, while measuring the amplitude of motion of the other drum that is probed with the red laser at a constant $V_{\rm{g},2}$. Figure~\ref{fgr:second}c shows the resonance peak splitting for different $V_{\rm g}$ applied to both membranes. Two distinct regimes are visible: one that corresponds to weak coupling at $V_{\rm g,2}=0$~V (and $J_{\rm{el}}=0$) and the other to strong coupling with the avoided frequency crossing visible at $V_{\rm g,2}=30$~V (and $J_{\rm{el}}\ne0$). We did not observe any change to the avoided crossing related to the change of laser intensity or its modulation amplitude (see SI 5). Also, resonance peaks disappear when the blue laser drive is focused on the unsuspended region of \ch{FePS3} (see SI 6). These observations show that neither a periodic heating from the laser beam, nor other parasitic electrical actuation mechanisms are responsible for the transfer of mechanical energy and strong coupling between the drums. Moreover, we observed the same behaviour in a test sample without a suspended channel connecting the two membranes, as shown in Fig.~\ref{fgr:second}d (see SI 7), thus providing additional evidence ruling out the possibility of strong tension-mediated direct mechanical coupling. We note that the non-zero amplitudes in weak coupling regime at $V_{\rm g,2}=0$~V in both Fig.~\ref{fgr:second}c and d indicate the presence of some other, much less pronounced, mechanisms of weak coupling (see SI 3). However, the evidence above suggests that the contribution of these mechanisms, that can couple the motion of two spatially separated \ch{FePS3} resonators under our experimental conditions, is negligible in comparison to the dominant mechanism that we propose in Fig.~\ref{fgr:second}b and eq~\ref{eq:F2ac} with $J\approx J_{\rm{el}}$. 

\paragraph{Strong Coupling Between Spatially Separated Nanodrums.}

\begin{figure*}
	\includegraphics[width=\linewidth]{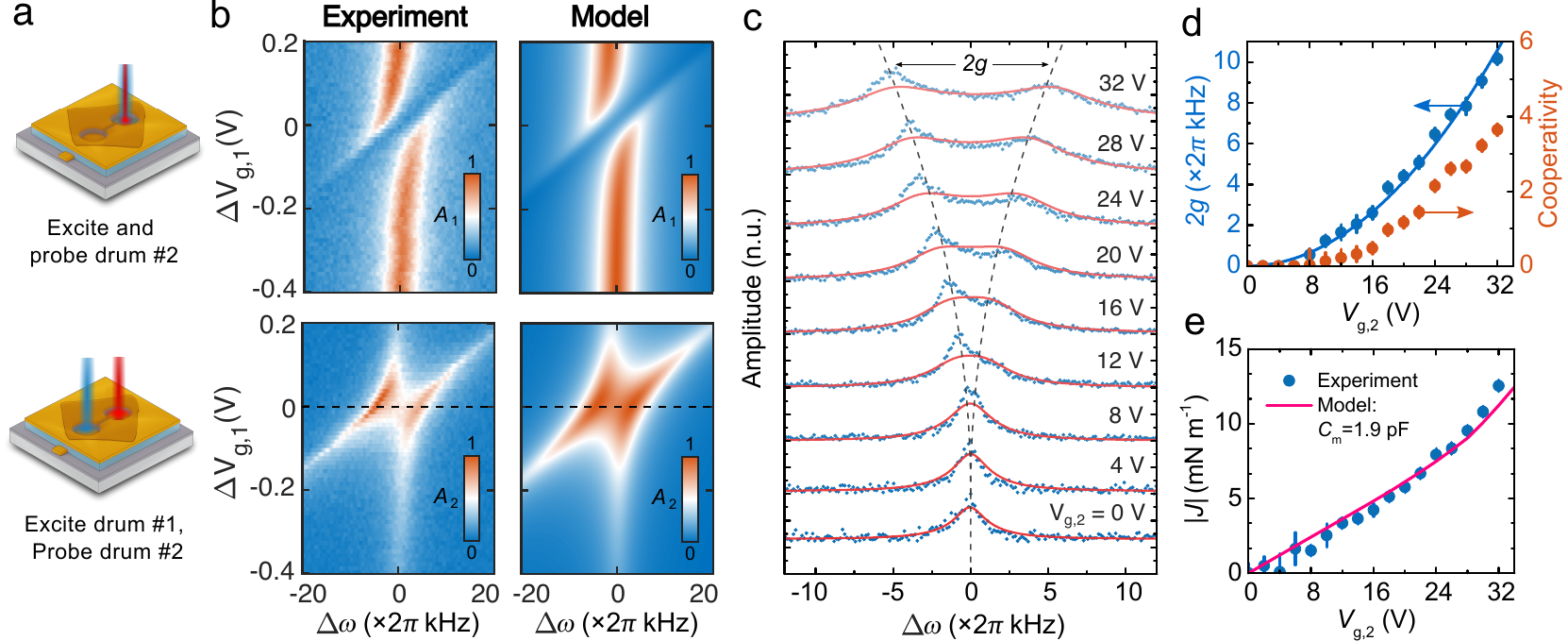}
	\caption{Comparison between the coupled oscillators model and experiments. (a) Schematic indication of the position of lasers for each row of data in (b). (b) Measured normalized amplitudes $A_{1,2}$ of the resonance peaks at $V_{\rm{g},1}=37.2$~V and $V_{\rm{g,2}}=30$~V compared with the model of eq~\ref{eq:system_j} and eq~\ref{eq:F2ac} (see SI 1-4). Dashed horizontal line examples the extraction of data shown in (c) at $\Delta \omega=\omega_{\rm d}-\omega_2$. (c) Amplitude $A_2$ of the resonance peak splitting of drum 2 at different $V_{\rm{g,2}}$. Filled blue dots - measured data. Solid red lines - fit to the model of eq~\ref{eq:amplitudes}. Dashed black line - positions of peak maxima used to extract $2g$. (d) Splitting $2g$ and cooperativity plotted against $V_{\rm g}$. Filled blue dots - measured $2g$ obtained from (c). Solid blue line - fit to a parabola used as a guide to the eye. Filled orange dots - cooperativity calculated from $2g$ and corresponding $\gamma_{1,2}$ from Fig.~\ref{fgr:first}g. (e) Measured and modeled coupling constant $J(V_{\rm g})$. Filled blue dots - $J$ extracted from the fit in (c). Solid magenta line - comparison to the model of eq~\ref{eq:F2ac}. Error bars in (d) and (e) are indicated with vertical colored lines.}
	\label{fgr:third}
\end{figure*}

Using the setup described above, we focus red and blue laser either on separate drums to excite one membrane and measure the motion of the other, or on the same drum to excite and measure a single one (see Fig.~\ref{fgr:third}a). We apply corresponding $V_{\rm{g}}$ to the drums in order to match their $\omega_{1,2}$ and measure the avoided crossing of the resonance frequencies in both configurations of lasers. By solving eq~\ref{eq:system_j}, we find amplitudes for the two configurations of lasers as \cite{Zanette}: 
\begin{equation}
     \begin{array}{ll}
        A_1=\frac{f_{\rm{d}}}{\gamma_1\omega_{\rm d}} \frac{\sqrt{\delta_2^2+1}}{\triangle},\\
        A_2=\frac{f_{\rm{d}}}{\gamma_1\gamma_2\omega_{\rm d}^2} \frac{|j|}{\triangle},
     \end{array}
     \label{eq:amplitudes}
\end{equation}
where $A_1$ is the oscillation amplitude with lasers on the same drum, $A_2$ the amplitude with lasers on different drums, $\triangle=\sqrt{\left(\Lambda+1-\delta_1\delta_2\right)^2+\left(\delta_1+\delta_2\right)^2}$, $\Lambda=\frac{j^2}{\gamma_1\gamma_2\omega_{\rm d}^2}$ the coupling strength coefficient, and $\delta_{1,2}=\frac{\omega_{1,2}^2-\omega_{\rm d}^2}{\gamma_{1,2}\omega_{\rm d}}$ the detuning. In Figure~\ref{fgr:third}b the measured amplitudes $A_{1,2}$ at $V_{\rm{g,2}}=30$~V are compared to simulations based on the continuum mechanics model (see SI 1) as well as eq~\ref{eq:system_j} and \ref{eq:F2ac}. The model is in good agreement with the experimental data (see SI 1-4). %describes the experiment well with the only exception of a small amplitude mismatch between the peaks in the measurement with separated lasers, that likely originates from a higher-order nonlinear effect.%

We now investigate the gate voltage dependence of strong coupling between the separated \ch{FePS3} membrane resonators. In Fig.~\ref{fgr:third}c we show the resonance peak splitting $2g$ with increasing $V_{\rm{g,2}}$. We extract $2g$ from peak maxima of the measured data in Fig.~\ref{fgr:third}c, which we plot together with the cooperativity calculated \cite{GrapheneMathew2016,GrapheneDistantLuo2018} as $\frac{(2g)^2}{\gamma_1\gamma_2}$. A strong coupling regime and an avoided crossing is reached when the figure of merit of the coupling, the cooperativity, is above $1$, which is achieved for $V_{\rm{g,2}}>16$~V.
We also fit $A_2$ of the same data set in Fig.~\ref{fgr:third}c to eq~\ref{eq:amplitudes} to extract $j$. Figure~\ref{fgr:third}e displays the measured coupling constant $J(V_{\rm g})$ (filled blue dots), compared to the electrical coupling model of eq~\ref{eq:F2ac} (solid magenta line). The model follows the experiment closely for $C_{\rm m}=1.9$ pF, reproducing both the quasi-linear part of the data at $V_{\rm{g,2}}<24$~V and the nonlinear part at $V_{\rm{g,2}}>24$~V that appears due to the deflection of the membrane $x_{\rm g}$ at larger $V_{\rm g}$.

\begin{figure}
	\includegraphics[scale=1.1]{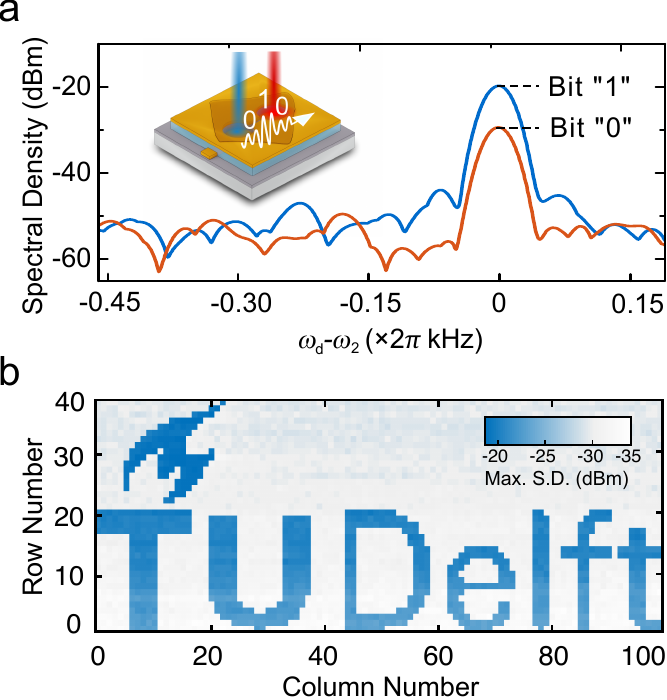}
	\caption{Information transfer between drums. (a) Measured spectral density near $\omega_2$ at $0$ and $1$ bit of amplitude modulated excitation. Inset: schematics of the experiment. (b) Map of the maximum of the spectral density showing a binary picture that was sent to drum 1 and received at drum 2 at a bit rate of $4$ bit/s.}
	\label{fgr:fourth}
\end{figure}

\paragraph{Amplitude Modulated Transmission of Information.}
In the strong coupling regime the excitation energy is transferred between the resonators. In Figure~\ref{fgr:fourth} we demonstrate that this channel of energy exchange can be amplitude-modulated to transfer binary data from one drum to another. We lock the gate voltage $V_{\rm g}$ of both drums and lock the excitation frequency at $\omega_{\rm d}$ as indicated with dashed lines in Fig.~\ref{fgr:second}c. Then, we modulate the drive power of the blue laser between $2.5$ and $5$ dBm with a step function, and thus the amplitude of excitation force $f_{\rm d}$ of drum 1, while measuring the motion of drum 2 using the red laser. The peak value in the measured spectral density corresponds to resonant motion of drum 2 at the excitation frequency $\omega_{\rm{d}}$ as shown in Fig.~\ref{fgr:fourth}a. The lower maximum of measured spectral peak density corresponds to a bit with a value 0, while the larger maximum to 1. Using this approach, we send a binary image to drum 1 and read it out on drum 2. The result is plotted in Fig.~\ref{fgr:fourth}b as a map of the maximum spectral density of the detected resonance peak on drum 2. The received picture is clearly distinguishable with no bits lost during the transfer.

\paragraph{Coupling Near the Antiferromagnetic N\'{e}el Temperature.}
%The  and dissipation rates of both resonators are expected to be strongly temperature dependent near $T_{\rm{N}}$ \cite{Siskins2020}%. 
\ch{FePS3} is an antiferromagnetic semiconductor at low temperature \cite{Haines2018, Evarestov2020} with a N\'{e}el temperature $T_{\rm{N}}\sim114$ K \cite{CvTakano2004, Lee2016}, where it exhibits a phase transition to a paramagnetic phase. The phase change in \ch{FePS3} is accompanied by a large anomaly in the thermal expansion coefficient that produces an accumulation of substantial tensile strain in the membrane \cite{Siskins2020} as it is cooled down from room temperature to $4$ K. As a consequence, at cryogenic temperatures membranes of \ch{FePS3}, even tens of nanometers thick \cite{Siskins2020}, have large quality factors of $2-6\times 10^4$ that are comparable to high-Q membranes made of strained monolayers of \ch{WSe2} and \ch{MoSe2} \cite{Morell2016}. In earlier works it was shown that the mechanical resonances of magnetic membranes can be sensitive to both the magnetic phase \cite{Siskins2020} and the magnetisation of that phase \cite{FaiMakJiang2020}. Therefore, when strongly coupled, small differences in magnetization can result in large differences in the resonance frequencies and the mechanical damping of the membranes and thus the coupling strength, making such coupled resonators very sensitive to small changes in the magnetic state of the material.

\begin{figure*}
\includegraphics[width=\linewidth]{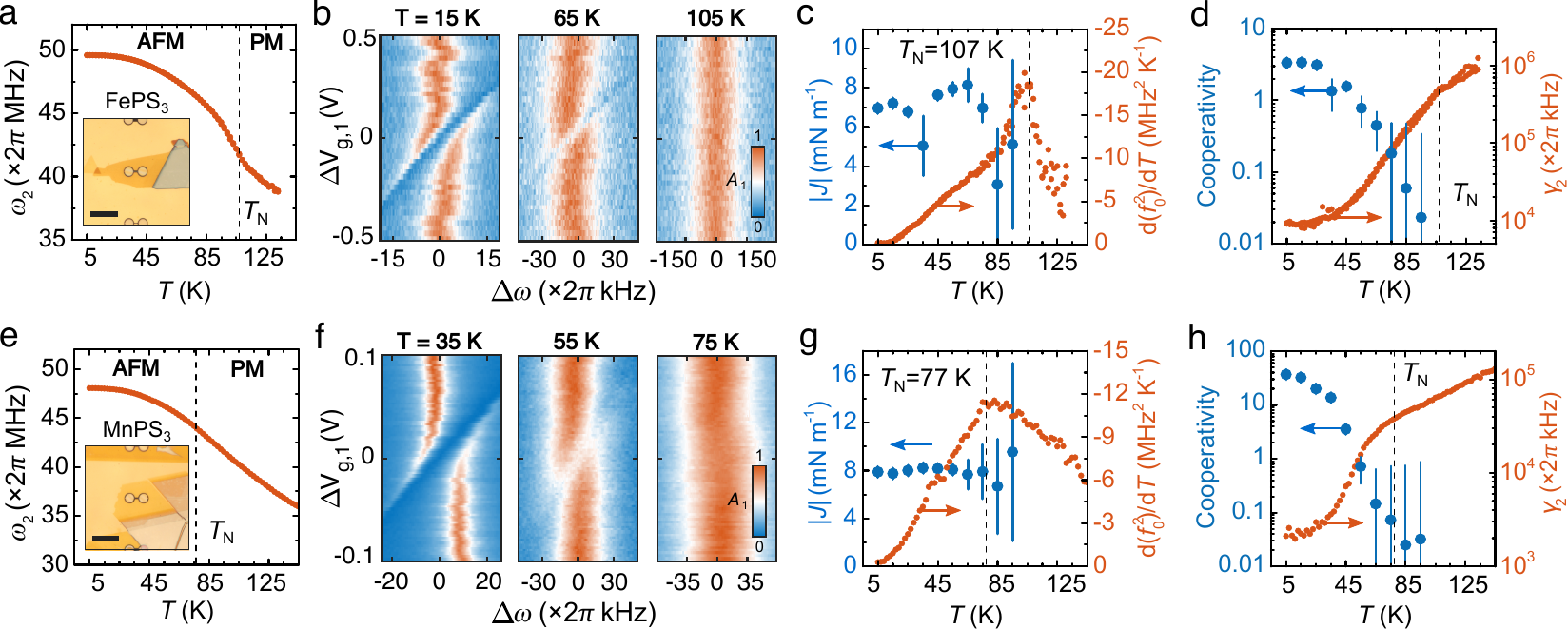}
	\caption{Temperature dependence of the coupling between antiferromagnetic membranes. (a) Resonance frequency $\omega_2$ of the FePS$_3$ drum as a function of temperature. Inset: Optical image of the FePS$_3$ sample. Thickness: $13.9\pm0.3$ nm. Scale bar: $18$ $\mu$m. (b) Normalized amplitude $A_1$ of the resonance peak splitting at $\Delta \omega=\omega_{\rm d}-\omega_2$ plotted for three different temperatures. (c) Filled blue dots - measured coupling constant $J$. Filled orange dots - $\frac{\mathrm{d}(f_0^2)}{\mathrm{d}T}$ of the data in (a). (d) Filled blue dots - cooperativity, filled orange dots - dissipation rate $\gamma_2$. The bottom panel (e-h) follows the same structure as the top panel (a-d) with the data shown for a \ch{MnPS3} sample. Inset of (e): Optical image of the \ch{MnPS3} sample. Thickness: $10.5\pm0.4$ nm. Scale bar: $18$ $\mu$m. Vertical dashed lines in all panels indicate the detected $T_{\rm N}$. Error bars in (c), (d), (g) and (h) are indicated with vertical blue lines.}
	\label{fgr:fifth}
\end{figure*}

In Fig.~\ref{fgr:fifth}a-d, we study a sample of FePS$_3$ to assess the temperature dependence of the coupling strength near the $T_{\rm N}$. Following the experimental organisation and analysis from above, we fix $V_{\rm{g,2}}$ of the drum 2 at $29$~V and measure the resonance frequency, coupling parameter and cooperativity as a function of temperature. As shown in Fig.~\ref{fgr:fifth}a, when the sample is heated up from $4$ to $135$ K, $\omega_{1,2}$ soften near the $T_{\rm{N}}$ of $107$ K. This also appears as a characteristic peak in $\frac{\mathrm{d}(f_0^2)}{\mathrm{d}T}=\frac{1}{4\pi^2}\frac{\mathrm{d}(\omega_2^2)}{\mathrm{d}T}$ in Fig.~\ref{fgr:fifth}c and originates from the anomaly in the specific heat of the material at $T_{\rm{N}}$ \cite{Siskins2020,CvTakano2004}. Interestingly, with the temperature approaching $T_{\rm{N}}$, the splitting of the resonance peak disappears, as shown in Fig.~\ref{fgr:fifth}b. However, as follows from eq~\ref{eq:F2ac}, $J_{\rm{el}}$ is not expected to have a strong temperature dependence or abrupt drop to zero near $T_{\rm{N}}$, which is also notable from Figure~\ref{fgr:fifth}c, where we plot the experimentally obtained $J(T)$. Instead, this switch from strong to weak coupling regime is related to a continuous decrease of the cooperativity due to an increasing $\gamma_{1,2}$ as it approaches the transition temperature, as shown in Fig.~\ref{fgr:fifth}d. This behavior of $\gamma_{1,2}(T)$ can be attributed to the increasing contribution of thermoelastic dissipation to the nanomechanical motion of drums near phase transitions \cite{Siskins2020}.

To support the hypothesis that the temperature dependent spectral changes are related to the temperature dependence of the dissipative terms in the equation of motion, we fabricated a sample of \ch{MnPS3} that exhibits the antiferromagnetic to paramagnetic phase transition at $T_{\rm{N}}\sim78$ K \cite{CvTakano2004}. In Fig.~\ref{fgr:fifth}e-h, we show the experimental data for \ch{MnPS3} that revealed a behavior similar to \ch{FePS3}. As shown in Fig.~\ref{fgr:fifth}e, $\omega_{2}$ softens near $T=77$ K which is close to $T_{\rm{N}}$. We observe the splitting disappearing next to $75\pm10$ K, as displayed in Fig.~\ref{fgr:fifth}f. As expected, $J$ does not show any systematic change near $T_{\rm{N}}$, which is depicted in Fig.~\ref{fgr:fifth}g. However, in Fig.~\ref{fgr:fifth}h the cooperativity has a sharper drop in value as the sample goes from strong to weak coupling regime with increasing temperature. This coincides with a broad kink in $\gamma_{1,2}$ that is visible near $T_{\rm{N}}$ of \ch{MnPS3}, providing evidence for the hypothesis.

\section{CONCLUSIONS}
In conclusion, we have demonstrated a mechanism that mediates strong coupling between spatially separated membranes made of antiferromagnetic materials \ch{FePS3} and \ch{MnPS3}. This coupling mechanism can be switched on and tuned by an electrostatic gate. In addition, the electromechanical transfer of energy can be amplitude modulated and is shown to be capable of performing bit-by-bit communication. This provides control advantages that can find the use in the development of new device concepts, such as nanomechanical logic gates \cite{LogicNEMSTsai2008,BitLogicMahboob2008, LogicMasmanidis2007} and hybrid systems combining magnetic mechanical oscillators and qubits \cite{SpinMEchanicsLee2017}. We have further shown that the magneto-mechanical properties of antiferromagnetic materials also can affect the coupling strength and cooperativity between the membranes next to the  phase transition. For example, we have shown that the increasing mechanical dissipation \cite{Siskins2020} near $T_{\rm N}$ of \ch{FePS3} and \ch{MnPS3} diminishes the cooperativity of such coupled membrane systems as $T$ approaches $T_{\rm N}$. Therefore, coupled NEMS made of magnetic membrane resonators can provide a deeper insight into coupling of magnetic properties to the nanomechanical motion. We also anticipate that in the future antiferromagnetic NEMS of this type can be useful to study more intricate magnetic phenomena, like a magnetostriction in ultrathin layers \cite{FaiMakJiang2020}, and the emission of spin currents by mechanical deformations - the piezospintronic effect \cite{PiezospintronicsUlloa2017}.

\section{EXPERIMENTAL METHODS}
\paragraph{Material Preparation.}
Crystal growth of MPS$_3$ (M = \ch{Mn}, \ch{Fe}) was performed following a solid state reaction inside a sealed evacuated quartz tube (pressure $\sim5\times10^{-5}$ mbar). For obtaining large crystals, \ch{I2} was used as a transport agent. The tube was placed in a three zone furnace with the material in the leftmost zone. The other two zones were heated up in $24$ hours from room temperature to $650$ $^{\circ}$C and kept at that temperature for one day. The leftmost side was then heated up to $700$ $^{\circ}$C in $3$ hours so that a temperature gradient of $700$ $^{\circ}$C/$650$ $^{\circ}$C/$675$ $^{\circ}$C was established. The temperature was kept constant for $28$ days and cooled down naturally. With this process crystals with a length up to several centimeters are obtained. Detailed description of the crystal growth and characterization can be found in earlier work \cite{Siskins2020}.  

\paragraph{Chip Fabrication.}
We first patterned the bottom Au electrodes with a Ti adhesive layer on top of \ch{Si}/\ch{SiO2} substrate using positive resist e-beam lithography. Following that, HSQ (FOX-12) spin-on-glass layers were spin-coated on top of the chip \cite{Davidovikj2017}. Subsequently, the chips were annealed in a furnace at $500$ $^{\circ}$C in Ar gas for 50 minutes. This cures the spin-on-glass, hardens it and improves the surface smoothness. This also provided an electrical insulation between subsequently deposited top layers and the bottom electrode. In the following step, we deposited the top \ch{Au} electrode layer with a Ti adhesive layer by evaporation. Using an additional \ch{Cr} layer as a hard mask, we defined circular cavities of $6$ $\mu$m in diameter and $x_{\rm c}=220$ nm in depth by e-beam lithography and reactive ion etching through the SOG. This also exposed the circular bottom \ch{Au} electrode of $5$ $\mu$m in diameter on the bottom of the cavity. The chip fabrication is described in detail in the previous work \cite{Davidovikj2017}. 

\paragraph{Sample Fabrication and Characterisation.}
Thin flakes of MPS$_3$ crystals were exfoliated and transferred on a prepatterned chip by an all-dry viscoelastic stamping method \cite{CastellanosGomez2014} directly after exfoliation. Atomic force microscopy (AFM) inspection and height profile scans were performed in tapping mode on a Bruker Dimension FastScan AFM. We used cantilevers with spring constants of $k=30-40$ Nm$^{-1}$. Error bars on reported thickness values were determined by measuring multiple profile scans of the same flake. Subsequently, samples were kept in an oxygen-free environment to avoid degradation. Following the cryogenic interferometry experiments, SEM imaging on some samples was performed using a FEI Nova NanoSEM 450 system.

\paragraph{Cryogenic Laser Interferometry.}
The sample was mounted on a piezoelectric $xy$ nanopositioning stage inside a dry cryostat chamber with optical access (Montana Instruments Cryostation s50). A modulated blue diode laser of $\lambda=405$ nm was used to optothermally excite the motion of the membrane. Membrane displacement detection was obtained by a laser interferometry technique using a focused He$-$Ne laser beam of $\lambda=632$ nm on the suspended membrane while recording the interfering reflections from the membrane and the \ch{Au} electrode underneath using a photodiode. The photodiode signal is processed by a vector network analyzer. All measurements were performed at incident laser powers of $P_{\rm{red}}\leq100$ $\mu$W and $P_{\rm{blue}}\leq1.5$ $\mu$W, unless stated otherwise. Laser spot size is on the order of $\sim1$ $\mu$m. It was checked for all membranes that resonance frequency changes due to laser heating were insignificant during the data acquisition. 

\section{ASSOCIATED CONTENT}
The Supporting Information is available free of charge at https://pubs.acs.org/doi/xxxxxx.

\section{AUTHOR INFORMATION}
\subsection{Author Contribution}
M.\v{S}., D.D., H.S.J.v.d.Z. and P.G.S. conceived the experiments. E.S. and M.\v{S}. performed the laser interferometry measurements. M.L. and E.S. fabricated and inspected the samples. S.M.-V. synthesized and characterized the \ch{FePS3} and \ch{MnPS3} crystals. M.\v{S}., E.S., and P.G.S. analyzed and modeled the experimental data. H.S.J.v.d.Z. and P.G.S. supervised the project. The paper was jointly written by all authors with a main contribution from M.\v{S}. All authors discussed the results and commented on the paper.
\subsection{Notes}
The authors declare no competing financial interest.

%\tableofcontents

\begin{acknowledgement}
M.\v{S}., M.L., H.S.J.v.d.Z. and P.G.S. acknowledge funding from the European Union's Horizon $2020$ research and innovation program under grant agreement number $881603$.  H.S.J.v.d.Z. and S.M.-V. thank COST Action MOLSPIN CA15128; S.M.-V. acknowledges the financial support from the European Union (ERC AdG Mol-2D 788222), the Spanish MICINN (MAT2017-89993-R co-financed by FEDER and Excellence Unit "Maria de Maeztu", CEX2019-000919-M) and the Generalitat Valenciana (Prometeo program and PO FEDER Program, ref. IDIFEDER/2018/061 and IDIFEDER/2020/063). We thank E. Coronado and S. Lodha for useful discussions and feedback on the manuscript.
\end{acknowledgement}

%\bibliography{bibl}% Produces the bibliography via BibTeX.
\providecommand{\latin}[1]{#1}
\makeatletter
\providecommand{\doi}
  {\begingroup\let\do\@makeother\dospecials
  \catcode`\{=1 \catcode`\}=2 \doi@aux}
\providecommand{\doi@aux}[1]{\endgroup\texttt{#1}}
\makeatother
\providecommand*\mcitethebibliography{\thebibliography}
\csname @ifundefined\endcsname{endmcitethebibliography}
  {\let\endmcitethebibliography\endthebibliography}{}

\end{document}